\documentclass[twocolumn,floats,showpacs,amsmath,amssymb,prb]{revtex4}

\usepackage{graphicx}

\begin{document}

\title{High-density amorphous ice: A path-integral simulation}
\author{Carlos P. Herrero}
\author{Rafael Ram\'{\i}rez}
\affiliation{Instituto de Ciencia de Materiales de Madrid,
         Consejo Superior de Investigaciones Cient\'{\i}ficas (CSIC),
         Campus de Cantoblanco, 28049 Madrid, Spain }
\date{\today}

\begin{abstract}
Structural and thermodynamic properties of high-density amorphous
(HDA) ice have been studied by path-integral molecular dynamics 
simulations in the isothermal-isobaric ensemble.
Interatomic interactions were modeled by using the effective q-TIP4P/F 
potential for flexible water.
Quantum nuclear motion is found to affect several observable
properties of the amorphous solid.
At low temperature ($T$ = 50 K) the molar volume of HDA ice is found
to increase by 6\%, and the intramolecular O--H distance rises by
1.4\% due to quantum motion.
Peaks in the radial distribution function of HDA ice are broadened
respect to their classical expectancy. 
The bulk modulus, $B$, is found to rise linearly with the pressure,
with a slope $\partial B/\partial P$ = 7.1.
Our results are compared with those derived earlier from classical and
path-integral simulations of HDA ice. We discuss similarities and 
discrepancies with those earlier simulations.
\end{abstract}



\maketitle

\section{Introduction}

Water is ubiquitous on Earth, and a detailed understanding of its
properties and behavior under different conditions is of crucial 
importance in several scientific fields.\cite{ei69,pe99,fr00,ro96} 
This does not only refer to thermodynamically stable water phases,
but also to various metastable phases obtained at ambient and extreme
conditions of temperature and pressure.
In particular, several forms of amorphous ice have been found and 
studied by both experimental\cite{mi84,mi85,he89,tu02b,ne06,st07} 
and theoretical\cite{ts99,gi05,ts05,lo06} methods, 
but some of their properties still lack a complete understanding.
This is mainly due to the peculiar structure of condensed phases of
water, where hydrogen bonds between adjacent molecules give rise to 
rather unusual properties of these phases. 

The atomic dynamics in amorphous solids cause the appearance of localized 
low-energy excitations, displaying appreciable
deviations from the situation of atomic nuclei harmonically
vibrating around their potential minima.\cite{cu87,el90}
Such deviations from harmonicity,
combined with the quantum character of atomic dynamics, is of
great importance in the characterization of amorphous materials.
 In this context, the classical papers by Phillips\cite{ph72}  
and Anderson {\em et al.}\cite{an72} opened a research line by 
modeling low-energy excitations in amorphous solids by two-level 
systems. In more recent years there have appeared several detailed 
descriptions of the low-energy motion in this type of materials,
beyond the standard tunneling model.\cite{ra98a}

 When one studies average structural properties of amorphous 
materials, an interesting question is whether quantum effects
can be noticeable in the presence of structural disorder. 
To be concrete, one may
ask whether the radial distribution function (RDF) obtained in
classical simulations is appreciably modified by considering
quantum atomic delocalization, which can be important mainly  
at low temperatures.
 Two factors appear to compete in the broadening of the RDF peaks 
at low temperature: structural disorder and quantum 
delocalization.\cite{he98}
As a first approach, one expects that for amorphous materials
with heavy atoms (small zero-point vibrational amplitudes),
structural disorder will broaden the peaks more than 
zero-point motion, and the opposite may occur for disordered
materials with light atoms.
For amorphous ice, one may suspect that the presence of hydrogen
will make quantum effects appreciable, even for strong structural 
disorder.

 Computer modeling of amorphous ice has been employed in recent 
years to obtain insight into its structural and dynamical 
properties.\cite{ga96,ok96,gi04,gi05,ts05,se09}
The beginning of computer simulations of condensed phases of water at 
an atomic level dates back more than 40 years,\cite{ba69,ra71}
and nowadays a large variety of empirical interatomic potentials can be 
found in the literature.\cite{ma01,ko04,jo05,ab05,pa06,mc09,go11}
Many of them assume a rigid geometry for the water molecule, whereas 
some others allow for molecular flexibility either with harmonic or 
anharmonic OH stretches. 
In recent years, simulations of water using \textit{ab initio} density 
functional theory (DFT) have been also carried out.\cite{ch03,fe06,mo08}
However, hydrogen bonds in condensed phases 
of water are difficult to describe with currently available energy
functionals, making that some properties are not accurately reproduced 
by DFT calculations.\cite{yo09}  Some progress in the description of 
van der Waals interactions in water within the DFT formalism 
has been recently made.\cite{wa11,ko11,ak11,pa12}

A shortcoming of {\em ab-initio} electronic-structure calculations is 
that they usually deal with atomic nuclei as classical particles, 
disregarding quantum effects like zero-point motion.
These effects may be accounted for using harmonic or quasiharmonic 
approximations for the nuclear motion, but  the precision of these
approaches is not readily estimated when large anharmonicities are present, 
as can be the case for light atoms like hydrogen in disordered materials.
To take into account the quantum character of atomic nuclei, 
the path-integral molecular dynamics (PIMD) 
approach has proved to be very useful,
since in this procedure the nuclear degrees of
freedom can be quantized in an efficient manner, thus including
both quantum and thermal fluctuations in many-body systems
at finite temperatures.\cite{gi88}
This computational technique is now well established as a tool to 
study problems in which anharmonic effects can be important.\cite{ce95}
Thus, a powerful approach could consist in combining DFT to determine the
electronic structure and path integrals to describe the quantum motion of
atomic nuclei.\cite{ch03,mo08} However, this procedure requires computer 
resources that would restrict enormously the number of state points that 
can be considered in actual calculations. 

Several forms of amorphous ice have been detected in recent years,
corresponding to different densities.\cite{gi05,lo06,ma05,bo06,lo11}
In the present paper we study high-density amorphous (HDA) ice by 
PIMD simulations at different 
pressures and temperatures, to analyze its structural and thermodynamic 
properties. 
Interatomic interactions are described by the flexible q-TIP4P/F model, 
which was recently developed and has been employed to carry out PIMD 
simulations of liquid\cite{ha09,ha11a,ra11} and solid\cite{ra10,he11,ha11b}
water.    Here we pose the question of how quantum motion of the
lightest atom can influence the structural properties of an amorphous
water phase, and in particular if this quantum motion is appreciable 
for the solid at different densities, i.e. under different
external pressures. This refers to the crystal volume and interatomic
distances, but also to the mechanical stability of the solid.
These questions have been addressed earlier for high- and low-density
amorphous ice.\cite{ga96,ur03}    In this context,
it is usually assumed that increasing quantum fluctuations 
enhances the exploration of the configuration space, but in certain
regimes an increase in quantum fluctuations can lead to dynamical arrest,
as found for glass formation.\cite{ma11}

 The paper is organized as follows. In Sec.\,II, we describe the
computational method and the model employed in our calculations. 
In Sec.~III we discuss the method employed to generate the simulation
cells of amorphous ice.
Our results are presented in Sec.~IV, dealing with molar volume, 
interatomic distances, kinetic energy, and bulk modulus of HDA ice.
 Sec.~V gives a comparison with earlier work, and Sec.~VI includes 
a summary of the main results.

\section{Computational Method}

We employ PIMD simulations to obtain several properties of amorphous 
ice at different temperatures and pressures.
This kind of simulations are based on an isomorphism between a 
quantum system and a classical one, obtained after a discretization 
of the quantum density matrix along cyclic paths.\cite{fe72,kl90}
This isomorphism corresponds to replacing each quantum particle by
a ring polymer consisting of $L$ (Trotter number) classical particles, 
joined by harmonic springs with temperature- and mass-dependent force
constant.
Details on this simulation procedure can be found elsewhere.\cite{gi88,ce95}
The dynamics used in this method is fictitious and does
not correspond to the real quantum dynamics of the considered
particles, but it helps to effectively sample the many-body configuration 
space, yielding precise results for the properties of the actual quantum 
system.
Another way to derive such properties could be the use of
Monte Carlo sampling, but we have found that this procedure requires for 
the present problem more computer resources than PIMD simulations.
An important advantage of the latter is that the computing 
codes can be more readily parallelized, which turns out to be a relevant 
factor for an efficient use of modern computer architectures.

Simulations of crystalline and amorphous ice were carried out here in the 
isothermal-isobaric $NPT$ ensemble ($N$, number of particles; 
$P$, pressure; $T$, temperature), which allows one to find the equilibrium 
volume of a solid at given pressure and temperature.
We have used effective algorithms for carrying out PIMD simulations 
in this statistical ensemble, similar to those described in the 
literature,\cite{ma96,tu98,tu02} and employed earlier in the study of
solid and liquid water by PIMD simulations.
We have considered temperatures between 50 K and 300 K, and pressures 
up to 8 GPa. 
Both negative (tension) and positive (compression) pressures have been 
employed in the simulations. For negative $P$ we considered amorphous
ice in the region of mechanical stability of the solid, down to
$P \sim -0.5$ GPa.
For comparison with results of PIMD simulations, some simulations of
HDA ice have been also performed in the classical limit, which 
is obtained in our path integral procedure by setting the Trotter 
number $L = 1$.

Our PIMD simulations were carried out on cells including 96 or
216 water molecules, which were generated from pressure treatment of
ice Ih and Ic supercells, respectively.
The former (96 molecules) corresponds to a $(3a, 2 \sqrt{3} a, 2c)$
supercell, where $a$ and $c$ are the standard hexagonal
lattice parameters of ice Ih, whereas the latter (216 molecules)
corresponds to a $3 \times 3 \times 3$ supercell of the cubic unit cell
of ice Ic.
The main purpose of taking these two ice types was to check the
influence of the starting crystalline ice on the properties of the
amorphous ice resulting under pressure. For all considered variables
(structural and thermodynamic), we found in both cases results that 
coincided with each other within statistical error bars of the
simulation procedure. 
To generate proton-disordered ice supercells (of Ih and Ic types) prior 
to amorphization, we employed a Monte Carlo procedure to impose
that each oxygen atom had two chemically bonded and two H-bonded
hydrogen atoms, and with a cell dipole moment close to
zero.\cite{bu98,he11}
For some particular conditions, we checked that both starting proton 
disorder and history of the amorphous supercell (amorphization
procedure) do not affect significantly the results presented below.
In particular, once formed the amorphous material from crystalline ice,
we checked that its properties are reversible upon increasing and 
decreasing temperature and/or pressure.

Interatomic interactions were modeled by the point charge,
flexible q-TIP4P/F model, developed to study liquid
water,\cite{ha09} and that was later employed to study various
properties of ice\cite{ra10,he11} and water clusters.\cite{go10}
Many of the empirical potentials used earlier for quantum
simulations of condensed phases of water treat H$_2$O molecules as rigid
bodies.\cite{he05,mi05,he06b} This turns out to be convenient for 
computational efficiency, but neglects the role of intramolecular
flexibility in the structure, dynamics, and thermodynamics of the
condensed water phases.\cite{ha09}
Moreover, the q-TIP4P/F potential takes into account the significant
anharmonicity of the O--H vibration in a water molecule by considering
anharmonic stretches, vs. the harmonic potentials employed in most
of the simulations that considered quantum effects in these water
phases.

Technical details on the PIMD simulations presented here are the same as 
those described and used in Refs.~\onlinecite{ra10,he11}.
The Trotter number $L$ has been taken proportional to the 
inverse temperature ($L \propto 1/T$), so that $L T$ = 6000~K,
which allows us to keep roughly a constant precision in the PIMD results
at the different temperatures under consideration. 
The time step $\Delta t$ employed for the calculation of interatomic forces
in the molecular dynamics procedure
was taken in the range between 0.1 and 0.3 fs, which was found to give
adequate convergence for the variables studied here.
For given conditions of pressure and temperature,
a typical simulation run consisted of $2 \times 10^5$ PIMD steps for 
system equilibration, followed by 
$10^6$ steps for the calculation of ensemble average properties.

\section{Preparation of amorphous ice}

In this Section we give details on the form in which we obtain the
disordered structures of HDA ice that are subsequently employed to 
characterize and study this amorphous phase from PIMD simulations.
In particular, we have obtained simulation cells of HDA ice by 
applying a hydrostatic pressure to ice Ih and ice Ic at several 
temperatures. This procedure allows us also to check the pressure at
which amorphization occurs, and to compare it with data (both
experimental and theoretical) reported in the literature.  
We note that HDA ice has been recently obtained at room temperature
from ice VII under rapid compression.\cite{ch11}

\begin{figure}
\vspace{-1.1cm}
\hspace{-0.5cm}
\includegraphics[width= 9cm]{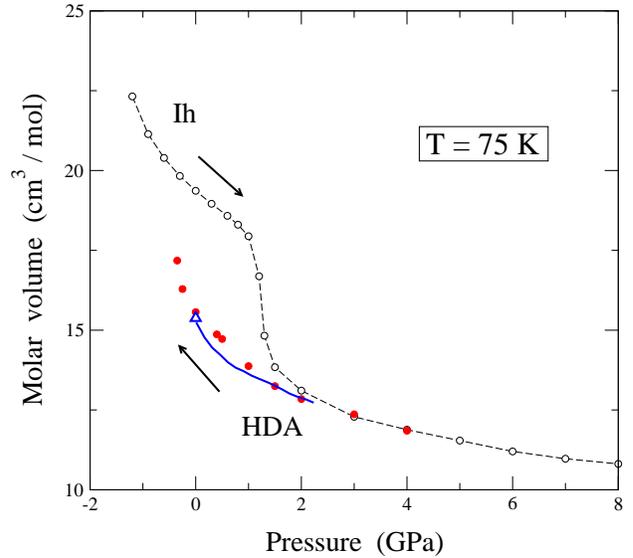}
\vspace{-0.8cm}
\caption{
Molar volume of ice as a function of pressure, as derived
from PIMD simulations at $T$ = 75 K.
Open circles represent results derived from simulations starting from
ice Ih.
Solid circles are data points obtained from simulations starting from
amorphous ice.
Error bars are in the order of the symbol size.
The dashed line is a guide to the eye.
An open triangle at $P$ = 0 indicates the volume measured by
Mishima {\em et al.}\cite{mi84}
The solid line was obtained from the pressure-density data displayed
in the review by Loerting and Giovambattista\cite{lo06}, taken from
Ref.~\onlinecite{ma04}.
}
\label{f1}
\end{figure}

In Fig.~1 we present the pressure dependence of the molar volume of 
ice at a temperature of 75 K, as derived from our PIMD simulations. 
Results shown as open circles correspond to simulations starting from
ice Ih, for both negative and positive pressures. We observe that in the
region between --1 GPa and 1 GPa the volume decreases smoothly as
pressure is increased, and at about 1.2 GPa it suffers a sudden
decrease which corresponds to ice amorphization. This value is close to
the spinodal pressure (limit of mechanical stability) obtained for 
ice Ih at this temperature in Ref.~\onlinecite{he11b} 
($P_s = 1.19 \pm 0.05$ GPa), and to the amorphization pressure obtained in
Ref.~\onlinecite{ok96} from classical molecular dynamics simulations.
From 1 to 2 GPa the volume decreases by 27\%, and at higher
pressures it continues decreasing to reach a value of 10.8 cm$^3$/mol
at $P$ = 8 GPa, to be compared with $v$ = 19.36 cm$^3$/mol found for 
ice Ih at atmospheric pressure.

We have carried out this procedure at several temperatures, and the
results are similar to those shown in Fig.~1. In particular, the
amorphization pressure changes slightly with temperature, and at
250 K we find 0.95 GPa. This was discussed in
Ref.~\onlinecite{he11b} in connection with the stability of ice Ih 
under pressure, and will not be repeated here. 
Apart from the pressure at which the solid amorphizes in the 
simulations, more precise values for the amorphization pressure can be 
obtained from the spinodal line of ice Ih, which gives the limit for
the mechanical stability of this solid phase. 
This pressure corresponds to the vanishing of the bulk modulus 
(divergence of the compressibility), and can be approached in computer
simulations at low temperatures ($T \lesssim 50$ K). 
At higher $T$ ice Ih amorphizes in the PIMD simulations before reaching 
the corresponding spinodal pressure, due to nucleation effects leading 
to the breakdown of the ice Ih structure.\cite{he11b}
We note that the formation of HDA ice from ice Ih has not been observed 
in the laboratory at temperatures so high as 250 K, but we obtain this 
transition at such temperatures and find an amorphous phase that remains 
metastable along our simulations (which in fact can only cover time scales
shorter than those of actual experiments).
HDA ice has been, however, obtained from ice VII at room 
temperature,\cite{ch11} as mentioned above.

Solid water amorphizes in an irreversible way, so that new PIMD
simulations at pressures lower than 1 GPa, starting from the amorphous
phase, do not recover the crystalline phase. 
This is shown in Fig.~1 as solid symbols. In these simulations the
pressure was gradually reduced down to negative pressures until
reaching the limit of mechanical stability of the material. 
At 75 K we could reach a pressure of --0.4 GPa, and at still more
negative pressures the amorphous solid broke down along the
simulations, transforming into the gas phase. This point will be discussed 
below in connection with the bulk modulus of HDA ice.
For comparison with our results we present in Fig.~1 the molar volume
obtained by Mishima {\em et al.}\cite{mi84} for HDA ice from x-ray
diffraction experiments at 80 K and atmospheric pressure.
Shown is also the volume-pressure curve derived from the data given in
the review by Loerting and Giovambattista,\cite{lo06} taken from
Ref.~\onlinecite{ma04} (solid line). 
We note that between $P$ = 0 and 1 GPa the molar volume of the amorphous 
material is clearly smaller than that of the crystal, but at negative
pressure the volume of the amorph increases fast, due to the
proximity of its metastability limit, and therefore approaches the
volume of ice Ih.

\begin{figure}
\vspace{-1.1cm}
\hspace{-0.5cm}
\includegraphics[width= 9cm]{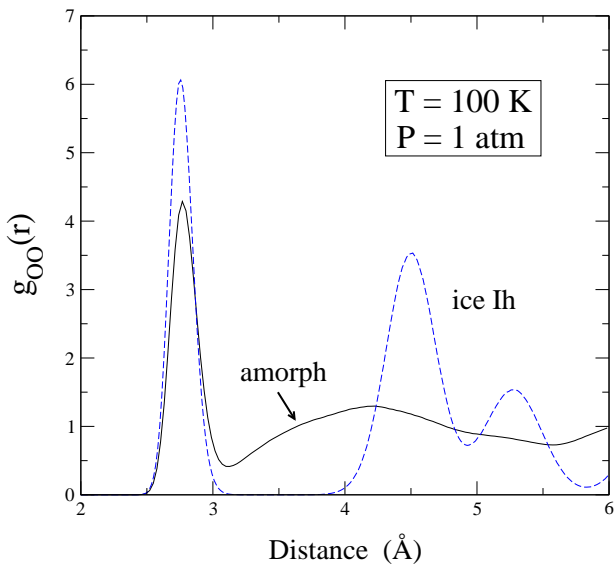}
\vspace{-0.8cm}
\caption{
Oxygen-oxygen radial distribution function of ice Ih (dashed line)
and HDA ice (solid line) at $T$ = 100 K and $P$ = 1 atm,
as derived from PIMD simulations.
}
\label{f2}
\end{figure}

For additional confirmation of the amorphous character of the solid
obtained after the cycle indicated by arrows in Fig.~ 1 (first
pressure increase up to 8 GPa, and then pressure reduction),
we display in Fig.~2 the O--O RDF for ice Ih and amorphous ice, as 
derived from PIMD simulations at $T$ = 100 K and atmospheric pressure.
Upon amorphization, the prominent peaks corresponding to the second and
third coordination sphere merge into one broad feature with a maximum
at about 4.1 \AA. 
The maximum of the first peak in the amorphous solid remains close to 
that of ice Ih, in spite of the appreciable volume reduction suffered
by the solid in the amorphization process. Note that the difference in
height of the first peaks in the crystalline and amorphous phases is
mainly due to the definition of radial distribution function, which is 
normalized by the density.\cite{ch87} 
In fact, we calculate $g(r)$ as
\begin{equation}
  g(r) =  \frac{d(r)}{D}  \; ,
\end{equation}
where $D$ is the mean density (atoms per unit volume) and $d(r)$ is 
the ``local density'' at distance $r$ from a reference atom:
\begin{equation}
  d(r) = \frac{N(r)}{4 \pi r^2 \Delta r} \; ,
\end{equation}
$N(r)$ being the number of atoms at distances between $r$ and 
$r + \Delta r$.

\section{Properties of high-density amorphous ice}

\subsection{Volume}

As shown above in Fig.~1 and discussed in Sect.~III, ice Ih suffers an
important reduction in volume upon amorphization at about 1.2 GPa.  
This volume decrease is associated to a softening of intermolecular O--H
bridges, accompanied by a reduction in the mean distance to
oxygen atoms in the second and third coordination shells (see Fig.~2).
After amorphization, the molar volume continues decreasing 
as pressure is raised, and we find at 75 K a reduction from 
13.11 cm$^3$/mol at $P$ = 2 GPa to 10.81 cm$^3$/mol at 8 GPa, which 
means a decrease of 18\% respect to the volume at 2 GPa. 

We emphasize that the volume of the amorph decreases smoothly in the whole
pressure region considered here, and we did not detect in this
region any other phase change as those reported in the literature.
In fact, Hemley {\em et al.}\cite{he89} observed a pressure-induced 
re-crystallization of HDA ice at about 4 GPa. Also, a slow transformation
of HDA to cubic ice on slow depressurization has been observed.\cite{jo07}
 In our simulations the amorph remains in its metastable state in the 
whole range of temperature and pressure studied here.  

\begin{figure}
\vspace{-1.1cm}
\hspace{-0.5cm}
\includegraphics[width= 9cm]{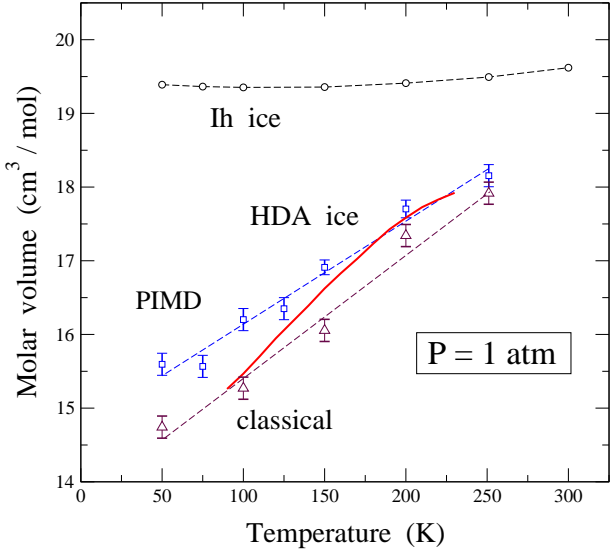}
\vspace{-0.8cm}
\caption{
Temperature dependence of the molar volume of ice at atmospheric
pressure as derived from PIMD simulations: Ih (circles) and
HDA ice (squares).
Results of classical simulations for HDA ice are displayed as triangles.
A solid line represents data obtained from classical molecular
dynamics simulations by Tse {\em et al.}\cite{ts05}
For ice Ih the error bars are smaller than the symbol size.
Lines are guides to the eye.
}
\label{f3}
\end{figure}

In Fig.~3 we show the molar volume of ice Ih (circles) and the HDA 
phase (squares) as a function of temperature at atmospheric pressure, 
as derived from our PIMD simulations.
At $P$ = 1 atm and $T$ = 75 K, after releasing the pressure applied for
amorphization, we find a molar volume $v$ = 15.57 cm$^3$/mol, which
corresponds to a density $\rho$ = 1.16 g/cm$^3$, in good agreement with 
the values given by Mishima {\em et al.} at zero pressure:
$\rho$ = 1.17 g/cm$^3$ in Ref.~\onlinecite{mi84} 
and 1.19 g/cm$^3$ in Ref.~\onlinecite{mi85}.
For comparison with the PIMD results, we also display in Fig.~3
classical data derived here with the q-TIP4P/F potential (triangles),
as well as those obtained by Tse {\em et al.}\cite{ts05} 
using the SPC/E potential.
At low temperature, the molar volume found by these authors is similar 
to that found here in the classical simulations, and it becomes
closer to the PIMD results as temperature rises.
Comparing our results with the q-TIP4P/F force field, we find at
50 K an increase in molar volume of about 0.85 cm$^3$/mol due to
nuclear quantum effects, which amounts to a 6\% of the classical
value.

In the temperature region up to 250 K we did not observe any transition
from HDA ice to a low-density amorphous phase with density
$\rho$ = 0.94 g/cm$^3$ (molar volume: 19.1 cm$^3$/mol), as experimentally 
observed at about 120--130 K,\cite{mi85,ts99,ne06,tu02b} and 
HDA ice remained as a metastable phase along our PIMD simulation
runs.  Something similar happens with the classical simulations reported in
Ref.~\onlinecite{ts05}. 
We believe that such a transition between amorphous phases is not captured
by the simulations due to the short time window that in fact can be
observed in the calculations, similarly to the difficulties found in this 
kind of simulations to directly obtain transitions between different 
crystalline phases.  This question could possibly be solved by performing  
direct coexistence simulations as those reported in the literature for 
liquid-solid transitions.\cite{ha09}

It is clear that the amorphous material has a higher density, or a smaller
molar volume than ice Ih, but it expands with temperature much faster than 
the crystalline solid.  At atmospheric pressure we find for the amorph 
a volume increase of 2.6 cm$^3$/mol in the temperature range from 
50 to 250 K.  In the same temperature region,
the volume of ice Ih is found to rise less than 0.3 cm$^3$/mol, in part
due to the negative thermal expansion around 100 K (not clearly observed 
at the scale of Fig.~3).
Thus, we find for the thermal expansion coefficient of HDA ice at 100 K:
$\alpha = 8.1 \times 10^{-4}$ K$^{-1}$.
This large volume expansion of the amorph, as compared with ice Ih, 
is indeed related with the lower bulk modulus of the amorphous
material at $P$ = 1 atm (see below).

\subsection{Interatomic distances}

\begin{figure}
\vspace{-1.1cm}
\hspace{-0.5cm}
\includegraphics[width= 9cm]{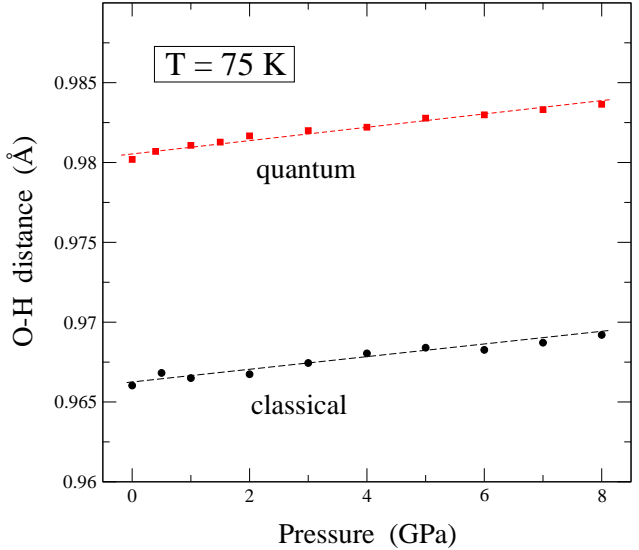}
\vspace{-0.8cm}
\caption{
Mean intramolecular O--H distance as a function of pressure for
amorphous ice at 75 K, as derived from PIMD (squares) and classical
(circles) simulations.
Error bars are in the order of the symbol size.
Lines are guides to the eye.
}
\label{f4}
\end{figure}

In this section we present results for interatomic distances
in amorphous ice between atoms in the same and adjacent molecules.
This can shed light on the structural changes suffered by the 
material when temperature and/or pressure are modified.
We first show in Fig.~4 the mean intramolecular O--H distance
as a function of pressure at a temperature $T$ = 75 K.
In connection with this figure, there are two results that should
be emphasized.
First, at a given pressure, we find that the O--H bond distance
increases appreciably due to nuclear quantum effects.  In fact,
a classical simulation at $T$ = 75 K and ambient pressure yields a mean
O--H distance of 0.966 \AA, to be compared with 0.980 \AA\ derived from
PIMD simulations, which means an increase of 1.4\% in the bond length
due to nuclear quantum motion.
This difference is much larger than the temperature-induced change in
$d$(O--H) at atmospheric pressure, which amounts to about 0.002 \AA\ in
the range from 50 to 250 K.
 The increase due to quantum motion is rather constant in the
whole pressure range studied here.

Another important result observed in Fig.~4 is that at a given
temperature the O--H distance increases as pressure is raised,
contrary to the usual contraction of atomic bonds for
increasing $P$.
This somewhat anomalous fact is due to the characteristic structure of
ice with hydrogen bonds connecting contiguous water molecules, 
which gives rise to an anticorrelation between the strength of molecular 
O--H bonds and intermolecular H bridges.\cite{li99}  
In fact, increasing the pressure causes a hardening of intermolecular H
bridges, with an associated weakening of intramolecular O--H bonds, and
a concomitant increase in the bond length.
 This weakening of intramolecular O--H bonds in ice for rising pressure 
has been observed experimentally and reported in the literature.\cite{ny06}

It is interesting to compare the O--H bond distance in amorphous ice
with that obtained for ice Ih in the same type of PIMD simulations.
At atmospheric pressure and 75 K, we find for ice Ih a mean distance
$d$(O--H) = 0.984 \AA, i.e. about 0.4\% longer than in HDA ice
at the same conditions.\cite{he11}
At the same temperature and $P$ = 1 GPa, close to the amorphization
pressure of ice Ih we found a distance difference of 0.6\%.
This means that the average O--H distance decreases upon amorphization
of the solid, which is consistent with a weakening of H bridges between 
adjacent molecules, causing a strengthening of the intramolecular bonds,
and therefore a shortening of the corresponding O--H distance.
These differences between O--H distances in Ih and HDA ice 
are consistent with changes in the stretching vibrational frequencies,
as observed from Raman scattering experiments. 
In fact, for HDA ice one observes at 80 K a broad Raman band with a 
maximum at $\approx$ 3200 cm$^{-1}$,\cite{sa06} to be compared with the 
largest feature appearing in the O--H stretching region of ice Ih at about 
3090 cm$^{-1}$.\cite{wa77} 
This hardening of the stretching vibrations upon amorphization
is consistent with the general trend found for water molecules, when the
intramolecular O--H distance decreases in different crystal 
surroundings:\cite{oj92}
$\Delta \omega$(O-H)/$\Delta d$(O-H) = --2.4 $\times 10^4$ cm$^{-1}$/\AA.
We note, for comparison, that Bellissent-Funel {\em et al.}\cite{be87}
found an intramolecular O--D distance of 0.97 \AA, from neutron
scattering experiments on high-density amorphous D$_2$O.

Another related aspect of the O--H distance is its temperature
dependence. For ice Ih at atmospheric pressure, this distance is known to  
decrease as temperature is raised, as a consequence of the hardening
of the bond for increasing temperature.  
This is in line with an enhancement of molecular motion for rising
$T$, which causes a weakening of the H bridges and an associated
enhancement of intramolecular bond strength.
Something similar is found for amorphous ice in our PIMD simulations,
where the average O--H distance decreases from 0.9804(2) 
to 0.9788(2) \AA\ when $T$ rises from 75 to 250 K.

\begin{figure}
\vspace{-1.1cm}
\hspace{-0.5cm}
\includegraphics[width= 9cm]{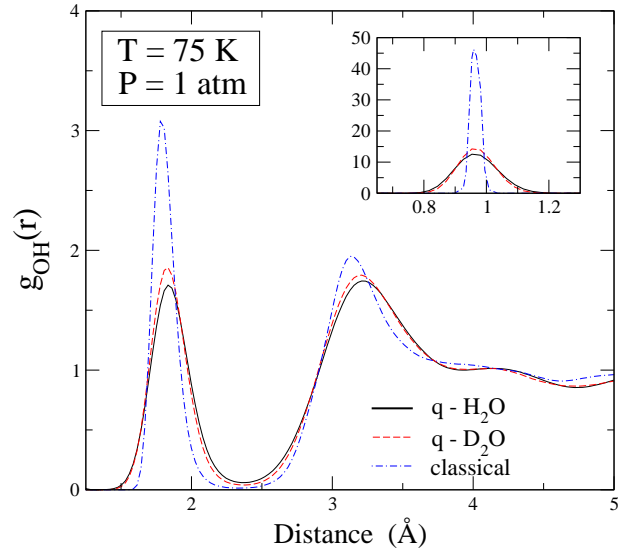}
\vspace{-0.8cm}
\caption{
Oxygen-hydrogen radial distribution function at 75 K and $P$ = 1 atm,
as derived
from quantum PIMD simulations for H$_2$O (solid line) and D$_2$O
(dashed line) amorphous ice, as well as from classical molecular
dynamics simulations (dashed-dotted line).
Inset: RDF in the region around 1 \AA, showing the peak corresponding to
intramolecular O--H bonds.
}
\label{f5}
\end{figure}

Given that nuclear quantum motion is appreciable in the mean O--H 
distance in water molecules in amorphous ice, it is expected that 
isotopic effects can be observed in the RDF.
In Fig.~5  we display the O--H radial distribution functions for
HDA ice, as derived from PIMD simulations for normal and
deuterated water, as well as from classical molecular dynamics
simulations. 
The RDF derived from classical simulations is similar to
that found earlier in this kind of simulations for HDA
ice.\cite{gi05}
For interatomic distances $r$ from  1.5 to 5 \AA, we observe in
Fig.~5 that quantum effects cause a broadening of the peaks. This 
is particularly observable in the peaks at about 1.8 \AA\ and 3.2 \AA.
The first peak in this RDF, corresponding to the intramolecular O--H
bonds, is much higher, and is displayed in the inset. 
For the classical model, it is about 15 times larger than the peak at 
1.8 \AA.
All peaks are found to broaden due to quantum effects, and their widths 
are larger for smaller isotopic mass. 
In fact, for the peak corresponding to intramolecular O--H bonds,
we obtained a full width at half maximum of 
0.05, 0.14, and 0.16 \AA, for classical ice, quantum D$_2$O, and 
quantum H$_2$O, respectively.
Note that in this respect the classical limit behaves as the large-mass
limit.

\begin{figure}
\vspace{-1.1cm}
\hspace{-0.5cm}
\includegraphics[width= 9cm]{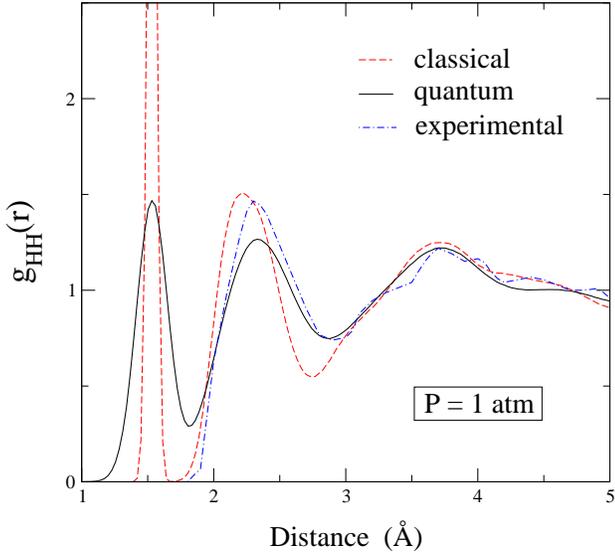}
\vspace{-0.8cm}
\caption{
Hydrogen-hydrogen radial distribution function for HDA ice at
$P$ = 1 atm.
Solid and dashed lines show results obtained from PIMD and
classical simulations, respectively, at $T$ = 75 K.
The dashed-dotted line was derived from neutron diffraction
experiments at 80 K.\cite{fi02b}
}
\vspace{0.6cm}
\label{f6}
\end{figure}

In Fig.~6 we show the hydrogen-hydrogen RDF of HDA ice as derived from
classical (dashed line) and PIMD simulations (solid line). 
As expected,
quantum motion of H broadens the peaks in the RDF with respect to the
classical result.
The peak corresponding to H--H pairs inside water molecules, at 1.53 \AA, 
has a height of 5.4 in the classical RDF, almost 4 times more than the 
quantum result of 1.45.  
In Fig.~6 we also present the H--H RDF obtained by 
Finney~{\em et al.}\cite{fi02} from neutron diffraction experiments 
at 80 K. Note that this curve does not include the intramolecular H--H
pairs in the original publication.
The next peak, corresponding to hydrogen pairs in adjacent (H-bonded)
molecules appears nearly at the same position in the RDF derived from
experiment and that obtained from PIMD simulations. However, the former 
is higher than the latter. This is the main difference between both
results, that in general are rather similar.

\begin{figure}
\vspace{-1.1cm}
\hspace{-0.5cm}
\includegraphics[width= 9cm]{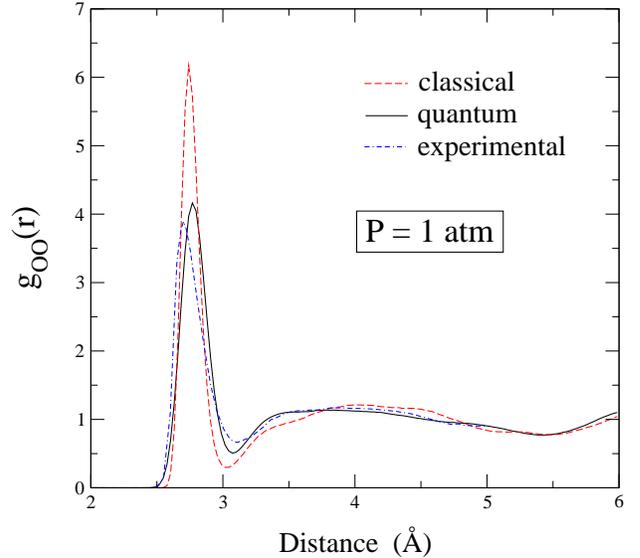}
\vspace{-0.8cm}
\caption{
Oxygen-oxygen radial distribution function for HDA ice at
$P$ = 1 atm.
Solid and dashed lines represent results derived from PIMD and
classical simulations, respectively, at $T$ = 75 K.
The dashed-dotted line was derived from neutron-scattering
experiments at 80 K.\cite{bo06}
}
\label{f7}
\end{figure}

Quantum effects are not only associated to hydrogen, due to its small 
mass, but are also observable in the oxygen-oxygen RDF of amorphous
ice, as displayed in Fig.~7. In this figure, the solid
line represents $g(r)$ derived from PIMD simulations at $T$ = 75 K and
atmospheric pressure, and the dashed line corresponds to the RDF derived 
from classical molecular dynamics simulations at the same conditions of
pressure and temperature. 
The classical result is similar to the oxygen-oxygen RDF derived in
Ref.~\onlinecite{se09} for several interatomic potentials.
For comparison, we also show in Fig.~7 the O--O RDF of HDA ice derived 
by Bowron {\em et al.} from neutron scattering experiments\cite{bo06}
(dashed-dotted line). As in the case of the O--H RDF, we observe a
broadening of the peaks when nuclear quantum motion is considered, as
a consequence of the associated atomic delocalization.
This is particularly observable for the first peak at about 2.8 \AA,
whose height decreases when quantum effects are taken into account. 
The result derived from PIMD simulations is closer to the experimental 
data than the classical one, but the peak derived from the quantum
simulations appears at $r$ = 2.77 \AA, a distance slightly larger than 
that corresponding to the maximum of the RDF derived from 
experiment ($r$ = 2.72 \AA).
We note, however, that the position of the peak derived from another 
neutron diffraction work\cite{fi02} is closer to that obtained in our 
simulations, but its height is somewhat smaller than that found here.
RDFs of HDA ice have been also derived from x-ray diffraction 
measurements.\cite{bo86}

It is also interesting to analyze the effect of pressure on the shape 
of the O--O RDF of amorphous ice, as it can give information on the 
molecular reorganization in the way to amorphous phases with still high
density.  With this purpose,
in Fig.~8  we present the O--O RDF at 75 K for different pressures:
atmospheric pressure along with $P$ = 1, 3, and 6 GPa.
We first observe that the peak at about 2.8 \AA, corresponding to the 
first coordination shell, moves to shorter distances as the pressure is
raised, in line with a decrease in the  O--O distance associated to
the corresponding volume reduction ($d V / d P < 0$).
It is also remarkable that the broad feature appearing in the RDF 
around 4 \AA\ (at atmospheric pressure) sharpens and moves to smaller 
distances for increasing pressure, indicating that water molecules in 
the second coordination shell come closer to those in the first shell.
This feature in the O--O RDF coincides with that observed in
very-high density amorphous (VHDA) ice.\cite{fi02b,lo06} 
Other features of the RDF appearing at larger $r$ also move to
shorter distances, as observed in the region between 5 and 6 \AA\ in
Fig.~8.
As shown above, we do not observe any clear discontinuity in the volume 
as pressure is increased, so that according to our results VHDA seems to 
appear as a high-pressure regime of HDA. 
Although in principle it is not evident
that the phase obtained by applying pressure to HDA ice is the same as
the VHDA obtained by isobaric heating,\cite{fi02b} our results are
in agreement with earlier simulations,\cite{ma05} favoring a continuous
transition from HDA to VHDA ice.
From an experimental point of view, there are indications\cite{lo06}  
pointing in the direction that structural differences between HDA and VHDA 
become less prominent as pressure increases.
In any case, there is some evidence against a first order-like
nature of the transition between HDA and VHDA, suggesting a
continuous character of the transition.\cite{lo11}

\begin{figure}
\vspace{-1.1cm}
\hspace{-0.5cm}
\includegraphics[width= 9cm]{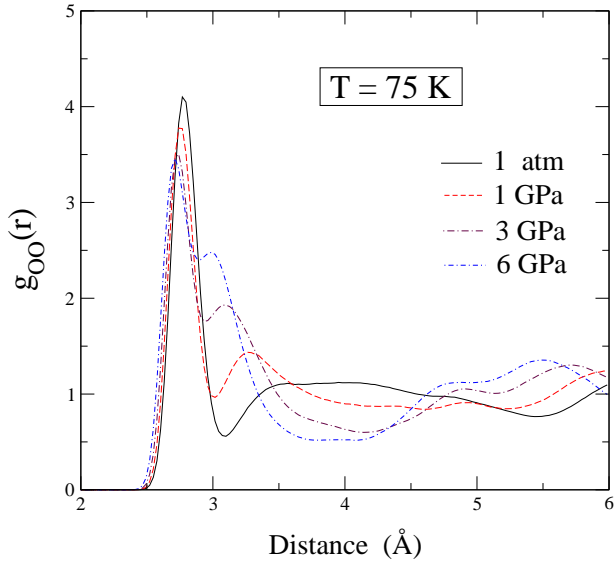}
\vspace{-0.8cm}
\caption{
Oxygen-oxygen radial distribution function for amorphous ice at
$T$ = 75 K and several pressures, as derived from PIMD simulations.
Different lines represent results for $P$ = 1 atm, and 1, 3, 6 GPa,
as indicated in the labels.
}
\label{f8}
\end{figure}

\subsection{Kinetic energy}

In this section we present the kinetic energy of atomic nuclei in
HDA ice, as derived from our PIMD simulations. 
The kinetic energy, $E_k$, of atomic nuclei depends on their mass and 
spacial delocalization, so that it can give information on
the environment and interatomic interactions seen by the considered
nuclei. We note that this does not occur for classical simulations, 
since in this case each degree of freedom contributes
to the kinetic energy by an amount that depends only on temperature,
i.e., $k_B T / 2$.
A typical quantum effect associated to the atomic motion in solids is
that the kinetic energy converges at low temperature to a value
related to zero-point motion, contrary to the classical result where
$E_k$ vanishes in the limit $T \to$ 0 K.
Path integral simulations allow us to obtain the kinetic energy
of the considered quantum particles.
For a particle with a certain mass at a given temperature, the larger 
the spread of the quantum paths, the smaller the kinetic energy,
in line with the expectancy that a larger quantum delocalization is 
associated with a reduction in the kinetic energy.\cite{gi88,he11}
We have calculated here the kinetic energy $E_k$ by using the so-called 
virial estimator, which has an associated statistical uncertainty 
appreciably lower than the potential energy of the system.\cite{he82,tu98}

\begin{figure}
\vspace{-1.1cm}
\hspace{-0.5cm}
\includegraphics[width= 9cm]{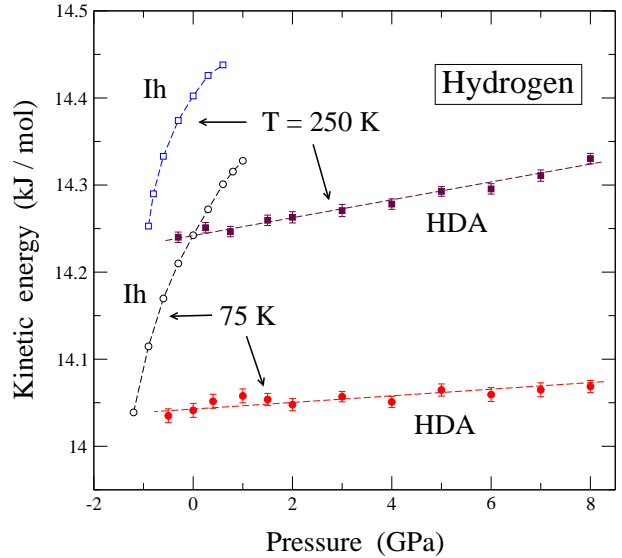}
\vspace{-0.8cm}
\caption{
Kinetic energy of hydrogen in ice Ih and amorphous ice as a function
of pressure at two temperatures: 75 K (squares) and 250 K (circles).
Open and solid symbols correspond to ice Ih and amorphous ice,
respectively.
Error bars for the crystalline phase are less than the symbol size.
Lines are guides to the eye.
}
\label{f9}
\end{figure}

 In Fig.~9 we display $E_k$ for hydrogen as a function of pressure 
at two temperatures: 75 and 250 K, as derived from our PIMD
simulations of amorphous ice (solid symbols). At each temperature, 
$E_k$ increases 
slowly as pressure rises, corresponding to an overall increase of 
vibrational frequencies.
For comparison, we also present results for hydrogen in ice Ih at the
same temperatures (open symbols). 
In this case, the kinetic energy increases with rising pressure
faster than for amorphous ice, in the whole region where ice Ih is
found to be mechanically stable.
This is a consequence of the larger quantum delocalization of hydrogen
in amorphous ice, as compared with ice Ih at the same temperature,
or equivalently, to the presence in the amorphous solid of modes
with lower vibrational frequency.
At atmospheric pressure, $E_k$ for HDA ice increases only by about
1.5\% from 75 to 250 K, reflecting the fact that at these temperatures 
most vibrational modes with large hydrogen contribution (i.e., libration
and stretching modes) are nearly in their ground state.
Moreover, changing the external pressure modifies even less
the kinetic energy of hydrogen. In fact, at 75 K it rises by about 
0.15\% in the range from atmospheric pressure to $P$ = 8 GPa.

\begin{figure}
\vspace{-1.1cm}
\hspace{-0.5cm}
\includegraphics[width= 9cm]{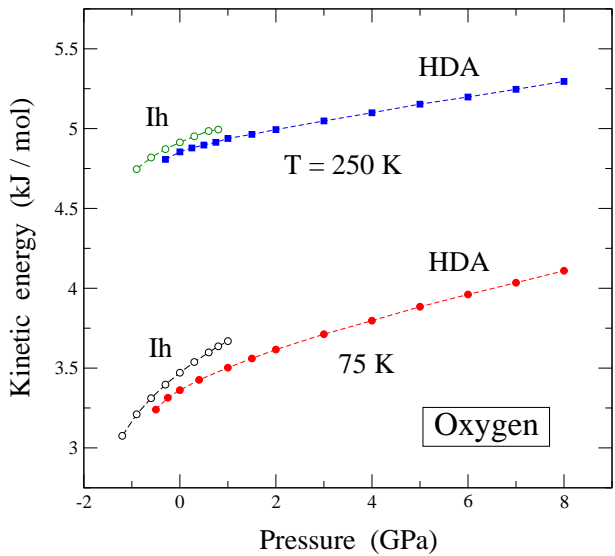}
\vspace{-0.8cm}
\caption{
Kinetic energy of oxygen in ice Ih and amorphous ice as a function
of pressure at two temperatures: 75 K (squares) and 250 K (circles).
Open and solid symbols correspond to ice Ih and amorphous ice,
respectively.
Error bars for the crystalline and amorphous solids are less than
the symbol size.
Lines are guides to the eye.
}
\label{f10}
\end{figure}

In Fig.~10 we show the pressure dependence of the kinetic energy of
oxygen in ice (HDA and Ih) at the same temperatures as those 
presented in Fig.~9 for hydrogen. 
We observe again that the kinetic energy rises with increasing pressure, 
and that at a given pressure, $E_k$ for oxygen in amorphous ice is 
smaller than in ice Ih.
At 75 K and atmospheric pressure the kinetic energy of oxygen results 
to be about 4 times smaller than that of hydrogen, as could be expected
from the larger mass of the former atom. At this temperature, $E_k$
for oxygen rises by about 22\% when a pressure of 8 GPa is applied.
This relative change is much larger than that found for hydrogen at
the same conditions (a 0.15\%).
For increasing temperature, we find also an important rise in $E_k$ for
oxygen, e.g., at $P$ = 1 atm we have a change of a 44\% from 75 to 
150~K, to be compared with a rise of 1.5\% in the case of hydrogen.  
This is indeed a consequence of the much larger mass of oxygen,
which contributes mainly to vibrational modes with low frequency, and 
therefore with excited states non-negligibly populated at these 
temperatures.

We note that the vertical scales in Figs.~9 and 10 are different, and what
seems to be a very large change in the kinetic energy of hydrogen when
comparing Ih and HDA ices, is in relative terms less than the change in
$E_k$ for oxygen.    In fact, 
at $P$ = 1 GPa and $T$ = 75 K (close to the amorphization pressure of
ice Ih), the kinetic energy of oxygen in HDA ice is 167 J/mol
lower than that in ice Ih, to be compared with a decrease of 270 J/mol
in $E_k$ for hydrogen.  In relative terms these differences amount to
a decrease of 4.5\% for oxygen vs. 1.9\% for hydrogen. 
This is in agreement with the fact that the structural changes associated
to ice amorphization affect mainly to low-frequency vibrational modes
(translational modes of the whole water molecule),
with a major relative contribution of oxygen to the kinetic energy.
The relative difference between kinetic energies in ice Ih and HDA
ice decreases as temperature is raised, as observed in Figs.~9 and 10 
at $T$ = 250 K, mainly because the nuclear motion becomes 
``more classical,'' and therefore $E_k$ is less sensitive to the environment 
and actual motion of the atomic nuclei under consideration. 

The kinetic energy of hydrogen and oxygen in liquid water
and ice Ih has been studied in detail earlier from PIMD
simulations, and compared with data derived from deep
inelastic neutron scattering in the case of hydrogen.\cite{ra11}
In that paper, a study of the contribution of different vibrational 
modes to the kinetic energy was presented.
For HDA ice we find here that, at a given temperature,
$E_k$ increases as pressure is raised for both hydrogen and
oxygen. This result is not obvious, since the O--H stretching
frequencies are known to decrease for increasing pressure
(their mode Gr\"uneisen parameter is negative\cite{ra12,pa12}). 
However, the overall contribution of the vibrational modes to $E_k$ 
increases with pressure, since the contribution of modes with positive
Gr\"uneisen parameter dominates, as analyzed elsewhere from a
quasi-harmonic approximation for crystalline ice phases.\cite{ra12}

\subsection{Bulk modulus}

The compressibility of ice displays peculiar properties associated
to the hydrogen-bond network.
For the crystalline phases of ice, and ice Ih in particular,
the compressibility is smaller than what one could suspect from the
large cavities present in its structure, which could be expected
to collapse under pressure before molecules could approach close enough
to repel each other.
This is in fact not the case, and for ice Ih the H bonds holding the
structure are known to be rather stable, as manifested by the
relatively high pressure necessary to break down the crystal.\cite{mi84}
Here we present results of PIMD simulations for the isothermal bulk modulus 
of HDA ice at different temperatures and pressures, and compare 
them with those derived for ice Ih.

The isothermal compressibility $\kappa$ of ice, or its inverse the bulk
modulus [$B = 1/\kappa = - V ( {\partial P} / {\partial V} )_T$] can be
directly derived from PIMD simulations in the isothermal-isobaric
ensemble.  In this ensemble, $B$
can be obtained from the mean-square fluctuations of the volume,
$\sigma_V^2 = \langle V^2 \rangle - \langle V \rangle^2$,
by using the expression\cite{la80,he08}
\begin{equation}
       B = \frac{k_B T \langle V \rangle}{\sigma_V^2}   \; ,
\label{bulkm}
\end{equation}
$k_B$ being  Boltzmann's constant.
This expression has been employed earlier to obtain the bulk modulus
of different types of solids from path-integral
simulations.\cite{he00c,he08,he11}

\begin{figure}
\vspace{-1.1cm}
\hspace{-0.5cm}
\includegraphics[width= 9cm]{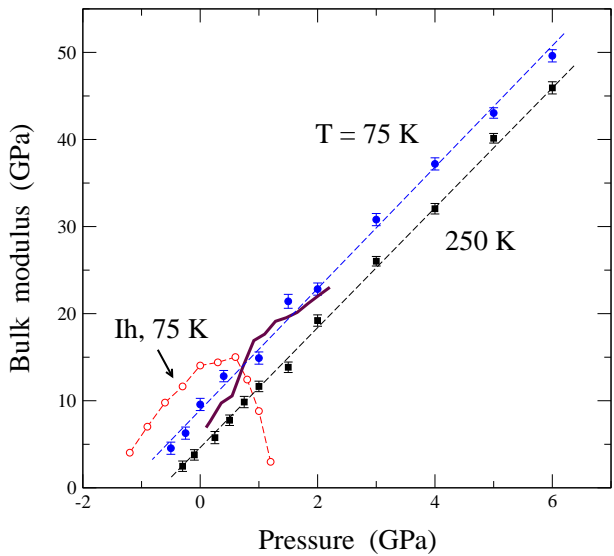}
\vspace{-0.8cm}
\caption{
Pressure dependence of the bulk modulus of amorphous ice at
$T$ = 75 K (solid circles) and 250 K (solid squares), as derived from
PIMD simulations.
Open circles correspond to ice Ih at 75 K.
Error bars for the crystalline phase are in the order of the symbol
size.
Dashed lines are guides to the eye.
The solid line was obtained by numerical differentiation from the
pressure-density data displayed in the review by Loerting and
Giovambattista,\cite{lo06} adapted from Ref.~\onlinecite{ma04}.
}
\label{f11}
\end{figure}

In Fig.~11 we present the bulk modulus of amorphous ice as a function
of pressure, as derived from our PIMD simulations at 75 K (solid circles) 
and 250 K (solid squares). For comparison, we also present results for 
ice Ih at 75 K (open circles).
The bulk modulus of amorphous ice is found to increase linearly
as a function of pressure in the range considered here. 
From linear fits to the data shown for amorphous ice in Fig.~11, we
find a slope $\partial B /\partial P$ = 7.1(2) and 7.0(2) at 75 and 
250 K, respectively, i.e. both values coincide within the precision of our
results.  This contrasts clearly with the pressure dependence of the 
bulk modulus found for ice Ih (open circles in Fig.~11). In fact,
for this crystalline phase of water one finds that at $P >$ 0.3~GPa, 
$B$ decreases as pressure is raised, until eventually reaching the limit
of mechanical stability of the phase (spinodal line) and the associated 
amorphization of the material. 
Thus, at pressures close to 1 GPa, the bulk modulus of
ice Ih is lower than that of the amorphous phase, contrary to the result
obtained at lower pressures.
For comparison, we also show in Fig.~11 the pressure dependence of the bulk
modulus of HDA ice at 80 K (solid line), derived by numerical differentiation 
of the pressure-density data given in Ref.~\onlinecite{lo06} (adapted 
from \onlinecite{ma04}).

At atmospheric pressure and 75 K, we find for amorphous ice 
$B$ = 9.6(4) GPa, to be compared with a value of $B$ = 14.0(3) GPa derived
for ice Ih at the same conditions.
This reflects an important increase in the compressibility of ice upon
amorphization, i.e. a lowering of the bulk modulus of the material by 
about a 30\%. 
Note that this reduction in the bulk modulus is similar to the relative 
volume change upon amorphization, as discussed in Sect.~III.

From our data in the temperature region between 75 and 250 K, $B$ 
extrapolates to zero at negative pressures in the order of --0.5 to --1 GPa, 
which gives the limit of mechanical stability of this amorphous  material,
where the solid breaks down giving rise to the gas phase.
As discussed elsewhere,\cite{sc95,he11b}
the possibility of studying a solid in metastable conditions,
close to a spinodal line is limited by the appearance of nucleation
events, which cause the breakdown of the solid.
For atomistic simulations such as those employed here, the probability of
those nucleation events at low temperatures is relatively low, and the
metastable range of the solid that can be explored is rather large.
In fact, we can go here to conditions near the limit of mechanical
stability at negative pressures (limit $B \to 0$).
As $T$ increases the probability of nucleation becomes higher, and the
accessible pressure range in the simulations is reduced.

\section{Comparison with earlier work}

 Data of earlier simulations of HDA ice have been already
presented in the previous section along with the results of our PIMD
simulations. Here we discuss and summarize the main similarities and
discrepancies between our results and those given in some earlier
works.

 Simulations of the amorphization of ice Ih using classical
molecular dynamics were carried out by Tse and Klein,\cite{ts87,ts90}
who employed the TIP4P intermolecular potential, and
found ice amorphization at pressures around 1.2--1.3 GPa at
temperatures between 80 and 100 K.
These values are close to the limit for mechanical stability
(spinodal pressure, $P_s$) of ice Ih obtained from PIMD simulations,
i.e., $P_s$ = 1.12 and 1.26 GPa at 50 and 100 K,
respectively.\cite{he11b}

Seidl {\em et al.}\cite{se09} carried out detailed classical
molecular dynamics simulations of HDA ice in the isothermal-isobaric
ensemble using several force fields. In particular, they studied
the glass-transition at a pressure of 0.3 GPa, and found some
indications of a glass-to-liquid transition at a temperature around
200 K, which could suggest that HDA ice is a proxy of an
ultraviscous high-density liquid. 
From our present results, we cannot find any evidence of
such a transition, and a detailed study of this point with PIMD
simulations would require at present enormous computational resources.

In connection with our work,
Tse {\em et al.}\cite{ts05} studied amorphous ice by classical molecular
dynamics simulations with the SPC/E potential. 
They presented O--O RDFs very similar to those
obtained in our classical simulations (shown in Fig.~7). 
These authors found a continuous and smooth increase in the molar volume 
of HDA ice as temperature is raised, without a sharp change 
indicating transformation from high-density to low-density amorphous phases. 
This is in line with the results of our PIMD simulations shown in Fig.~3.
In particular, they found at 100 K a molar volume of 15.4 cm$^3$/mol,
close to our classical result (triangles in Fig.~3) and smaller than
the value derived from our PIMD simulations with the q-TIP4P/F potential 
($v$ = 16.2 cm$^3$/mol). 
At higher temperatures the results by Tse {\em et al.} become closer
to our PIMD results, and are therefore somewhat higher than our
classical data. This can be a consequence of the differences between 
the force fields employed in both works.

Earlier studies of path integral simulations of amorphous ice are
scarce.
Gai {\em et al.}\cite{ga96} studied structural properties of HDA ice
at 77 K by path-integral Monte Carlo simulations, using the SPC/E
potential model, which treats the water molecules as rigid bodies.
Their simulations were carried out in the constant volume, canonical
($NVT$) ensemble, and dealt with both H$_2$O and D$_2$O ice.
The O--H RDF obtained by these authors for D$_2$O amorphous ice is
similar to that obtained here, and presented in Fig.~5.
In particular, they found well-defined peaks for the first and second
coordination shells at about 1.8 and 3.3 \AA, respectively.
However, for H$_2$O ice they found a RDF very different to that found
here, in which the peaks at 1.8 \AA\ (intermolecular O--H bridges)
and 3.3 \AA\ were missing, and had been replaced by a broad feature 
extending from about 2.5 to 4 \AA. These authors argued that the
hydrogen bonding network that is present for D$_2$O either disappears 
or is totally mixed with the second nearest-neighbor shell. 

Apart from the constant volume employed by Gai {\em et al.}\cite{ga96}
in their simulations vs the constant pressure employed in ours, 
the main difference between
both kinds of calculations seems to be the interatomic potential:
rigid molecules in Ref.~\onlinecite{ga96} vs flexible molecules in our
calculations. We do not find, however, a direct explanation why the use of
rigid molecules should cause a so strong difference between the structures
of H$_2$O and D$_2$O amorphous ice, as suggested by the results of
Gai {\em et al.}

\section{Summary}

 PIMD simulations provide us with a suitable tool to analyze effects 
of nuclear quantum motion in amorphous ice at finite temperatures.
Here, we have presented results of PIMD simulations of HDA ice 
in the isothermal-isobaric ensemble at different pressures and temperatures.
This kind of simulations have allowed us to obtain structural and
thermodynamic properties of this metastable material in a large region of
pressures, including tensile stresses ($P < 0$). 

The HDA ice studied here was obtained computationally by pressure-induced 
amorphization of cubic and hexagonal ice (Ih and Ic).
We observe an important reduction of the volume upon amorphization at a
pressure of about 1.2 GPa. 
The resulting ice at atmospheric pressure and 100 K is 19\% denser 
than its crystalline precursor,
but it is found to be softer, in the sense that its compressibility and
thermal expansion coefficient are clearly larger. 
At $P$ = 1 atm and $T$ = 75 K, the compressibility $\kappa$ of HDA
ice is about 50\% larger than that of ice Ih, and the thermal
expansion coefficient $\alpha$ for the amorphous solid is found to be
$9 \times 10^{-4}$ K$^{-1}$, whereas it is negative for ice Ih at
75 K.

We have assessed the importance of quantum effects by comparing results
obtained from PIMD simulations with those obtained from classical
simulations. 
Structural variables are found to change when nuclear quantum motion is
considered, especially at low temperatures. Thus, the crystal volume, 
interatomic distances, and
radial distribution functions suffer appreciable modifications in the 
range of temperature and pressure considered here. 
At 50 K the molar volume of HDA ice is found to rise by 0.85 cm$^3$/mol
(a 6\% of the classical value), and the intramolecular O--H distance 
increases by 1.4\% due to quantum motion.

The zero-point vibrational motion of atomic nuclei is large enough 
to also change appreciably structural observables of the amorphous
solid, such as the radial distribution function at low temperatures.
In fact, from PIMD simulations we observe a broadening of the peaks 
in the RDFs, as compared with classical molecular dynamics
simulations. For different isotopes we also observe a change in the 
RDFs of HDA ice. In particular, the width of the peaks in the 
O--H RDF is found to depend on the hydrogen isotope under
consideration, but the general features of this RDF are basically the
same for both H$_2$O and D$_2$O, contrary to earlier path-integral
Monte Carlo results.\cite{ga96}

At a given temperature, the kinetic energy of both hydrogen and oxygen is
found to increase for rising pressure. This increase is, however, slower
than that obtained in ice Ih.
Such an increase is associated to the rise in vibrational zero-point
energy and an overall increase in the vibrational frequencies of
the solid. However, the intramolecular O--H distance is found to
increase as pressure is raised, with a decrease in the frequency of
the corresponding stretching modes.
In HDA ice the bulk modulus is found to increase linearly as a 
function of pressure, in the whole region studied here. At 75 K we 
find $\partial B / \partial P$ = 7.1. 

Although quantitative values found by using the q-TIP4P/F potential 
can change by employing other interatomic potentials, 
the main conclusions obtained here can hardly depend on the potential 
employed in the simulations.
 Quantum simulations similar to those presented here can
give information on the atom delocalization and anharmonic effects
in other kinds of amorphous ice. 
An extension of this work could consist in studying amorphous ice at 
still higher pressures, where very-high density amorphous ice could be
characterized, including nuclear quantum effects.

\begin{acknowledgments}
This work was supported by Ministerio de Ciencia e Innovaci\'on (Spain)
through Grant FIS2009-12721-C04-04 and by Comunidad Aut\'onoma de Madrid 
through Program MODELICO-CM/S2009ESP-1691.
\end{acknowledgments}

\end{document}